\documentclass[]{ceurart}
\usepackage{dsfont} 
\usepackage{stmaryrd} 
\usepackage{amsmath} 
\usepackage{amssymb}
\usepackage{newpxtext}
\usepackage{booktabs}
\usepackage{graphicx}
\usepackage{tikz} 
\usepackage{robust-externalize}
\robExtConfigure{enable fallback to manual mode} 

\usetikzlibrary {arrows.meta, automata, positioning}

\makeatletter
\newcommand{\ie}{\emph{i.e.~}\@ifnextchar.{\!\@gobble}{}}
\newcommand{\eg}{\emph{e.g.~}\@ifnextchar.{\!\@gobble}{}}
\newcommand{\etc}{\emph{etc.~}\@ifnextchar.{\!\@gobble}{}}
\makeatother
\newcommand\scalemath[2]{\scalebox{#1}{\mbox{\ensuremath{\displaystyle #2}}}}

\newcommand{\AtL}{\mathrm{Atoms}(\Lambda)}
\newcommand{\RRd}{\mathds{R}^{d}}
\newcommand{\PP}{{\mathcal P}}

\usepackage{algorithm,algpseudocode}
\algnewcommand{\Input}[1]{
  \State \textbf{Input:}
  \Statex \hspace*{\algorithmicindent}\parbox[t]{.8\linewidth}{\raggedright #1}
}
\algnewcommand{\Output}[1]{
  \State \textbf{Output:}
  \Statex \hspace*{\algorithmicindent}\parbox[t]{.8\linewidth}{\raggedright #1}
}
\algnewcommand{\Initialize}[1]{
  \State \textbf{Initialize:}
  \Statex \hspace*{\algorithmicindent}\parbox[t]{.8\linewidth}{\raggedright #1}
}
\algrenewcommand{\Return}{\State\algorithmicreturn~}
\usepackage{listings}
\begin{document}

\copyrightyear{2025}
\copyrightclause{Copyright for this paper by its authors.
  Use permitted under Creative Commons License Attribution 4.0
  International (CC BY 4.0).}

\conference{OVERLAY 2025: 7th Workshop on Artificial Intelligence and Formal Verification, Logic, Automata, and Synthesis,
October 26th, Bologna, Italy}

\title{Passive Learning of Lattice Automata from Recurrent Neural Networks}

\author[]{Jaouhar Slimi}[
email=jaouhar.slimi@cea.fr
]
\author[]{Tristan {Le Gall}}[
email=tristan.le-gall@cea.fr
]
\author[]{Augustin Lemesle}[
email=augustin.lemesle@cea.fr
]
\address[]{Université Paris-Saclay, CEA, List, France}

\begin{abstract}
We present a passive automata learning algorithm that can extract automata from recurrent networks with very large or even infinite alphabets. Our method combines overapproximations from the field of Abstract Interpretation and passive automata learning from the field of Grammatical Inference. We evaluate our algorithm by first comparing it with the state-of-the-art automata extraction algorithm from Recurrent Neural Networks trained on Tomita grammars. Then, we extend these experiments to regular languages with infinite alphabets, which we propose as a novel benchmark.
\end{abstract}

\maketitle

\section{Introduction}
\label{sec:intro}
Recurrent Neural Networks (RNNs) are a class of Deep Neural Networks conceived primarily to process sequential data. Their linear complexity in space and time, combined with their efficiency, leads to their adoption in various applications from Time Series Forecasting \cite{lin2023segrnnsegmentrecurrentneural} to policy modelling in reinforcement learning environments \cite{Schrittwieser_2020}. RNNs have been extensively studied from the lens of automata and formal language theory to model and analyze their behavior \cite{Jacobsson2005RuleEF, Bollig2022ASO}.
Concretely, automata learning from RNNs seeks to map (or approximate) its semantics to some kind of automaton.

Automata learning methods are either based on active learning, where a form of feedback is received from the System Under Learning (SUL), which is the RNN in our case, or passive learning, where we only require a set of execution traces. Our work focuses on the latter approach. Our goal is to construct a symbolic automaton; more precisely, a Lattice Automaton, from a set of traces of an RNN. Lattice Automata are similar to Deterministic Finite Automata~(DFA), however, they can recognize languages over an \emph{infinite alphabet}, such as the inputs of an RNN, which are usually in the form of floating-point vectors. Therefore, we can be closer to its real behavior, and give a surrogate model of an RNN trained to recognize a series of real-world inputs (e.g. language modelling tasks such as sentiment analysis \cite{Johnson2016SupervisedAS}). 

\paragraph{Outline:} We first briefly present a few key notions needed to understand our work. Then, we present our technique and some experimental results. We conclude with the advantages and limitations of this approach, and our future work.

\paragraph{Related work:}
In recent years, there has been a renewed interest in learning automata from RNNs, for interpretability as well as for verification purposes. 
Active learning techniques are mainly based on extensions of the L* algorithm for learning regular languages. Multiple works have been proposed \cite{ANGLUIN198787, Mayr2021PropertyCW, Mukardin2022LearningFS} to extend L* for extracting automata from RNNs.

On the other hand, passive learning techniques rely on the hypothesis that in RNNs, semantically similar intermediate neurons (also called hidden states) tend to cluster together~\cite{muškardin2023relationshiprnnhiddenstate}. 
Multiple techniques relying on clustering were proposed  for learning DFA~\cite{adax, Wang2018ACS}, Weighted Finite Automata~(WFA)~\cite{Wei2023WeightedAE}, Hidden Markov Models~(HMM)~\cite{Song2023LUNAAM}, and Probabilistic Finite Automata~(PFA) \cite{Dong2019AnalyzingRN}. The type of automata influences the complexity of the solution and is mainly motivated by the application; for instance, DFAs are well-suited to deterministic RNN behaviors, whereas WFAs and PFA are better alternatives to capture their probabilistic dynamics.
A different approach by \cite{merrill2022extractingfiniteautomatarnns} relies on the classical Gold algorithm \cite{GOLD1967447} by creating a prefix tree automaton~(PTA) from the execution traces, then merging its states to yield a DFA. The authors associate RNNs hidden states values to each state of the PTA and then use that during the merge algorithm by only merging states that have at most a distance $d$ between their associated hidden states. In this paper, we adapt the latter method to learn automata over an infinite alphabet.

\section{Key Notions}
\label{sec:keynotions}

\paragraph{Elman Networks} A category of simple RNN architectures,  defined by a function $F_{\theta}(x_{t}, h_{t-1})$ that computes the hidden state $h_t$, followed by a classifier function $G_{\phi}(h_t)$ that computes the output $y_t$. $F_{\theta}$ and $G_{\phi}$ are non linear functions parameterized by their weights $\theta$ and $\phi$. For readability purposes, we assume that both the input vector $x_{t}$  and the hidden state $h_{t-1}$ are vectors in $\mathds{R}^{d}$, at any time step $t = 1,2 \dots n$ \: \footnote{if the dimensions are not the same, we can always resize them to the highest dimension}. In our paper, for simplicity, we consider only binary classification networks, \ie $y_t \in \mathds{B}$.
The computation carried out by the RNN after processing an input vector $x_t$ is then in the following form:
$(x_{t}, h_{t-1}) \xrightarrow{\quad F_{\theta} \quad} h_{t} \xrightarrow{\quad G_{\phi} \quad} y_{t}$.

\paragraph{Abstract Lattice} Let us consider finite sequences on $\RRd$ (which is our infinite alphabet). We want to abstract $(\PP(\RRd), \subseteq)$ using a lattice $(\Lambda, \sqsubseteq)$, which is defined as follows:
\begin{itemize}

\item A relation $\sqsubseteq$ on a set $\Lambda$ is a \emph{partial order} if: 
\begin{enumerate}
    \item $\forall \lambda \in \Lambda, \lambda \sqsubseteq \lambda$
    \item $\forall \lambda_1,\lambda_2 \in \Lambda, (\lambda_1 \sqsubseteq \lambda_2) \wedge (\lambda_2 \sqsubseteq \lambda_1) \Rightarrow \lambda_1 = \lambda_2$
    \item $\forall \lambda_1,\lambda_2,\lambda_3 \in \Lambda, (\lambda_1 \sqsubseteq \lambda_2) \wedge (\lambda_2 \sqsubseteq \lambda_3) \Rightarrow \lambda_1 \sqsubseteq \lambda_3$
\end{enumerate}

\item $(\Lambda,\sqsubseteq)$ is a \emph{complete lattice} if:
\begin{enumerate}
    \item $\sqsubseteq$ is a \emph{partial order} on a set $\Lambda$
    \item any subset of $\Omega \subseteq \Lambda$ has a greatest lower bound $\sqcap \Omega$ (\ie the set $\{ y \in \Lambda \, | \, \forall x \in \Omega, x \sqsupseteq y\}$ has a greatest element), and a least upper bound $\sqcup \Omega$ (\ie the set $\{ y \in \Lambda \, | \, \forall x \in \Omega, x \sqsubseteq y\}$ has a smallest element)
\end{enumerate}

\item Then we define:
\begin{itemize}
    \item[$\cdot$] $\bot = \sqcup \emptyset = \sqcap \Lambda$
    \item[$\cdot$] $\top = \sqcap \emptyset = \sqcup \Lambda$
    \item[$\cdot$] $\AtL$ as the set of the minimal elements of $\Lambda \setminus  \bot$; in other words, $a \in \Lambda$ is an atom if $\forall \lambda \in \lambda, \lambda \sqsubseteq a \Rightarrow \lambda = a \vee \lambda = \bot$
\end{itemize}

\end{itemize}

\paragraph{Atomistic Lattice} A lattice $(\Lambda,\sqsubseteq)$ is \emph{atomistic} if: $\forall \lambda \in \Lambda: \lambda = \sqcup \{a \in \AtL \, | \, a \sqsubseteq \lambda \}$. Therefore, we can define a finite partition of $\AtL$ either as we did previously, or as a function $\Pi: \{1 \dots n\} \to \Lambda$ satisfying the following property: $\forall a \in \AtL$ there exists a unique $i \in \{1 \dots n\} $ such that $a \sqsubseteq \Pi(i)$. In this paper, and for the sake of simplicity, $(\Lambda, \sqsubseteq)$ is defined by the box abstraction.

\paragraph{The Box Abstraction} Any set $\Omega \subseteq \RRd$ can be abstracted by a box ($d$-uple of intervals) $\alpha(\Omega) = \langle \omega_1 \dots \omega_d \rangle$, such that each $\omega_i$ is an interval bounding the projection of $\Omega$ on the dimension $i$. It is a classical abstract domain, and there is a Galois connection~\cite{CC77} between $(\PP(\RRd), \subseteq)$ and $(\Lambda, \sqsubseteq)$. This lattice is atomistic: its atoms are the singletons $ \langle \llbracket x_1,x_1 \rrbracket \dots \llbracket x_d,x_d \rrbracket \rangle$ for any vector $\langle x_1 \dots x_d\rangle \in \RRd$, thus the isomorphism between $\AtL$ and $\RRd$. Because of this isomorphism, we identify any $x\in \RRd$ with $\alpha(x) \in \Lambda$. That is why, in the definition of the language recognized by a Lattice Automaton, we wrote the condition "$x_i \sqsubseteq \lambda_i$", while strictly speaking, the condition should be "$\alpha(x_i) \sqsubseteq \lambda_i$" since $x_i \in \RRd$.

\paragraph{Partition} A finite partition on $\AtL$ is a function $\Pi: \{1 \dots n\} \rightarrow \PP(\AtL)$ such that $\bigcup_{i=1}^n \Pi(i) = \AtL$ and $i \neq j \Rightarrow \Pi(i) \cap \Pi(j) = \emptyset$. Since $\Lambda$ is atomistic, for any $i \in \{1 \dots n\}$, we can identify the set of atoms $\Pi(i)$ with $\sqcup \{a | a \in \Pi(i) \}$ which defines the "maximal element" (also noted $\Pi(i)$ by abuse of notation) of this class. We have the property that classes of the partition are stable for the operation $\sqcup$ : $\forall \lambda_1, \lambda_2: \lambda_1 \sqsubseteq \Pi(i) \wedge  \lambda_2 \sqsubseteq \Pi(i) \Rightarrow \lambda_1 \sqcup \lambda_2 \sqsubseteq \Pi(i)$. Partitions are needed to properly define Lattice Automata, see \cite{Gall2007LatticeAA} for details.

\paragraph{Lattice Automata}
Lattice Automata are similar to DFA, but with transitions labelled by the elements of the atomistic lattice $\Lambda$ rather than elements of a finite alphabet $\Sigma$. Formally, a Lattice Automaton is a tuple $\mathcal{A} = \langle \Lambda, \Pi, Q, \mathbb{I}, \mathbb{F}, \delta, \Gamma \rangle$ such that:\\
\begin{minipage}{0.5\textwidth}
\begin{itemize}
    \item $\Lambda$ is an atomistic lattice
    \item $\Pi$ is a partition of $\AtL$
    \item $Q$ is a finite set of states
\end{itemize}
\end{minipage}
\begin{minipage}{0.5\textwidth}
\begin{itemize}
    \item $\mathbb{I} \subseteq Q$ are the initial states
    \item $\mathbb{F} \subseteq Q$ are the final states
    \item $\delta \subseteq Q \times \Lambda \times Q$ is the set of transitions
    \item $\Gamma: Q \xrightarrow{} \Lambda$ is the hidden states function 
\end{itemize}
\end{minipage}

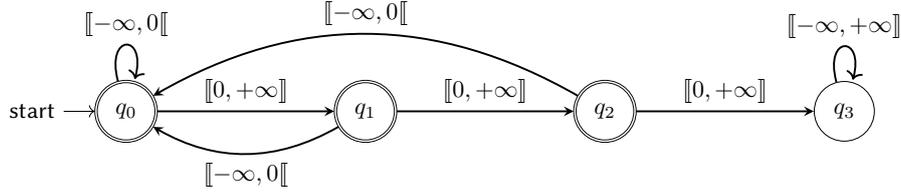
\begin{figure}[!h]
\centering
\scalebox{0.9}{
\begin{tikzpicture} [node distance = 3.5cm, on grid, auto]
\node (q0) [state, initial, accepting, initial text = {start}] {$q_0$};
\node (q1) [state, accepting, right = of q0] {$q_1$};
\node (q2) [state, accepting, right = of q1] {$q_2$};
\node (q3) [state, right = of q2] {$q_3$};

\path [-stealth, thick]
    (q0) edge [loop above] node [above] {$\llbracket -\infty, 0 \llbracket$} (q0)
    (q0) edge node[above] {$\llbracket 0, +\infty \rrbracket$} (q1)
    (q1) edge node[above] {$\llbracket 0, +\infty \rrbracket$} (q2)
    (q2) edge node[above] {$\llbracket 0, +\infty \rrbracket$} (q3)
    (q1) edge [bend left] node[below] {$\llbracket -\infty, 0 \llbracket$} (q0)
    (q2) edge [bend right] node[above] {$\llbracket -\infty, 0 \llbracket$} (q0)
    (q3) edge [loop above] node[above] {$\llbracket -\infty, +\infty \rrbracket$} (q3);
\end{tikzpicture}
}
\caption{Example of a Lattice Automaton. $\mathbb{I} = \{q_0\}, \; \mathbb{F} = \{q_0,q_1,q_2\}$.}
\label{fig:ila_example}
\end{figure}

We also require the following two properties to ensure that the number of transitions is finite, even if the alphabet is infinite: 
\begin{enumerate}
    \item For any transition $(q,\lambda,q') \in \delta$, there is $i \in \{1 \dots n\}$, such that: $\forall a \in \AtL, a \sqsubseteq \lambda \Rightarrow a \in \Pi(i)$. In other words, a transition shall not mix atoms belonging to different classes of the partition. Thus, we can define a function $\Pi^{-1} : \Lambda \to \{1 \dots n\}$ that associates a label $\lambda$ of a transition to its class of the partition.
    \item For any couple of states $(q,q')$ and any class of the partition $i\in\{1 \dots n\}$, there is at most one transition $(q,\lambda,q) \in \delta$ such that $\Pi^{-1}(\lambda) =i $.
\end{enumerate}

In this work, since in our examples and experiments $(\Lambda,\sqsubseteq)$ we rely on the box abstraction, we will simply call them Interval Lattice Automata (ILA).

\paragraph{Language recognition}
 Like a DFA, an ILA recognizes a language, \ie a set of words on the input alphabet $\AtL \equiv \RRd$. A word $x_1 \dots x_k$ ($x_i \in \RRd$ for all $i$) is accepted by $\mathcal{A}$ if there is a sequence $q_0 \xrightarrow{\lambda_1} q_1 \xrightarrow{\lambda_2} \dots \xrightarrow{\lambda_k} q_k$ such that:
\begin{itemize}
    \item $\forall \: i=\{1 \dots k\}, (q_{i-1},\lambda_i,q_i) \in \delta$ and $x_i \sqsubseteq \lambda_i$
    \item $q_0 \in \mathbb{I}$ and $q_k \in \mathbb{F}$
\end{itemize}

\paragraph{Representation of the hidden states} The function $\Gamma$ does not play any role in the definition of the language recognized by an ILA. Its purpose is to associate any state of the automaton with a set of hidden states of the RNN, abstracted by an element of $\Lambda$. Unlike the elements labelling the transitions, we do not require $\Gamma(q)$ to belong to a single class of the partition $\Pi$, it can be anything (including $\top$). We included this function in the definition of the automaton since it has a role in the learning algorithm.

\paragraph{Adding transitions} Algorithms on ILA are similar to the ones on DFA. However, one must ensure that the two properties stated above remain true. For example, when we want to add a transition $(q, \lambda, q')$ to an ILA, we must first check if there already exists a transition between the two states $q$ and $q'$, labelled by $\lambda'$ belonging to the same class as $\lambda$. If so, that transition is replaced by $\lambda \sqcup \lambda'$, as written in Algorithm~\ref{alg:add_trans}.

\begin{algorithm}[!h]
\caption{Add Transition}
\label{alg:add_trans}
\begin{algorithmic}[1]
\Input{An automaton $\mathcal{A} = \langle \Lambda, \Pi, Q, \mathbb{I}, \mathbb{F}, \delta, \Gamma \rangle$ and $(q,\lambda,q') \in Q \times \Lambda \times Q$ such that $\exists i \in \{1\dots n\} \lambda \sqsubseteq \Pi(i)$}
\Output{$\mathcal{A}_m = \langle \Lambda, \Pi, Q_m, \mathbb{I}_m, \mathbb{F}_m, \delta_m, \Gamma_m \rangle$ }
\Initialize{\strut
$Q_m \gets Q $ \\ 
$\mathbb{I}_m \gets \mathbb{I} $ \\
$\mathbb{F}_m \gets \mathbb{F} $ \\
$\delta_m \gets \delta$ \\
$\Gamma_m \gets \Gamma$
}

\State $j \gets \Pi^{-1}(\lambda)$
\If{$\exists (q, \lambda', q')  \in \delta_m$ such that $ \Pi^{-1}(\lambda')=j$}
\State $\delta_m \gets \delta_m \setminus \{(q,\lambda',q')\} \cup \{(q, \lambda \sqcup \lambda',q')\}$;
\Else
\State $\delta_m \gets \delta_m  \cup \{(q, \lambda,q')\}$;
\EndIf

\Return $\mathcal{A}_m$;
\end{algorithmic}
\end{algorithm}

\section{Learning Lattice Automata from RNN execution traces}
\label{sec:learning}

The proposed algorithm is an extension of Gold's algorithm for the case of ILA. We first create an Interval Prefix Tree Automaton (IPTA) from a set of traces of the RNN. Then, we launch the merging phase in which we merge states according to a \emph{similarity score} inspired by \cite{merrill2022extractingfiniteautomatarnns}. 

\begin{figure}[!h]
\scalebox{0.9}{
\begin{tikzpicture}[shorten >=1.0pt,node distance=3.2cm, on grid, thick,>={Stealth[round]}]

  \node[state,initial, accepting] (q_0) {$q_0$};
  \node[state, accepting] (q_1) [above right=of q_0] {$q_1$};
  \node[state, accepting] (q_2) [below right=of q_1] {$q_{2}$};
  \node[state, accepting] (q_20) [right=of q_1] {$q_{20}$};
  \node[state, accepting] (q_3) [right=of q_2] {$q_3$};
  \node[state, accepting] (q_4) [right=of q_3] {$q_4$};
  \node[] (q_4+) [right=of q_4] {$\dots$};
  \node[state] (q_21) [right=of q_20] {$q_{21}$};
  \node[] (q_21+) [right=of q_21] {$\dots$};

  \path[->] (q_0) edge node [below right]{$\scalemath{0.9}{\llbracket 1.4, 3.39 \rrbracket}$} (q_1)
            (q_1) edge node [above right]{$\scalemath{0.9}{\llbracket -3.2, -1.07 \rrbracket}$} (q_2)
            (q_2) edge node [below left, pos=0.8]{$\scalemath{0.9}{\llbracket 1.08, 7.91 \rrbracket}$} (q_3)
            (q_3) edge node [below left, pos=1.0]{$\scalemath{0.9}{\llbracket -7.06, -3.45 \rrbracket}$} (q_4)
            (q_4) edge node [below left, pos=0.8]{$\scalemath{0.9}{\llbracket 2.1, 9.03 \rrbracket}$} (q_4+)
            (q_1) edge node [above right, pos=0.05]{$\scalemath{0.9}{\llbracket 2.29, 3.56 \rrbracket}$} (q_20)
            (q_20) edge node [above left, pos=0.9]{$\scalemath{0.9}{\llbracket 2.06, 3.14 \rrbracket}$} (q_21)
            (q_21) edge node [above left, pos=0.9]{$\scalemath{0.9}{\llbracket -33.2, -0.51 \rrbracket}$} (q_21+);
\end{tikzpicture}
}
\caption{Example of an IPTA created given the set S from a RNN trained on language 4 of Tomita 2.0}
\label{fig:example_ipta}
\end{figure}
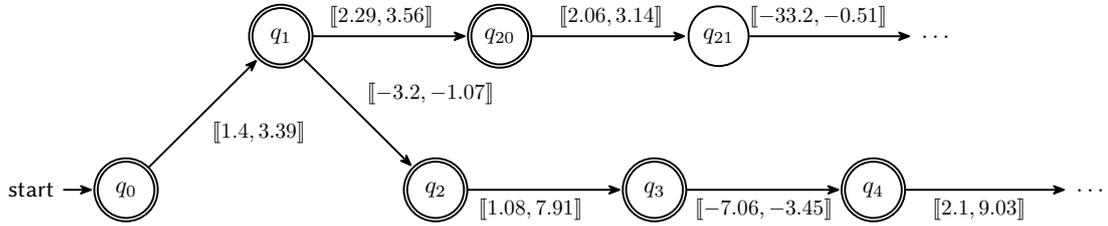

\paragraph{Building the Interval Prefix Tree Automaton (IPTA)} Let us consider a finite set of RNN execution traces $S=\{ s_i = \left(x_1^{(i)}, h_1^{(i)}, y_1^{(i)} \right) \dots  \left(x_{k_i}^{(i)}, h_{k_i}^{(i)}, y_{k_i}^{(i)} \right)  |\: k_i \geq 1\} $, where sequences $s_i$ are of varying lengths $k_i$. We build the IPTA starting with an automaton with a single initial state $q_0$ and applying the function $\Call{Add Sequence}{\mathcal{A},s}$ for every sequence $s\in S$. The function $\Call{Add Sequence}{\mathcal{A},s}$ takes a sequence $s = \left(x_{1}, h_{1}, y_{1} \right),\left(x_{2}, h_{2}, y_{2} \right)  \dots  \left(x_{k}, h_{k}, y_{k} \right)$ and adds transitions and states to the automaton $\mathcal{A}$ in the following way:

We consider the initial state $q_0$ and the first triplet $\left(x_1, h_1, y_1 \right)$ of the sequence:
\begin{itemize}
    \item If there exists already in $\mathcal{A}$ a state $q_1$ and a transition $(q_0,\lambda_1,q_1)$ such that $\Pi^{-1}(\lambda_1)=\Pi^{-1}(x_1)$, then we modify the transition $(q_0,\lambda_1',q_1)$; with $\lambda' = \lambda \sqcup \llbracket x_1, x_1 \rrbracket$, and the $\Gamma$ function to $\Gamma(q_1) \gets \Gamma(q_1) \sqcup \llbracket h_1, h_1 \rrbracket$; moreover, $q_1$ becomes a final state if $y_1=1$
    \item otherwise, we create a new state $q_1$ and a transition $(q_0,x_1,q_1)$, with $\Gamma(q_1) \gets h_1$ and $q_1$ being a final state if $y_1=1$
\end{itemize}
The process is then repeated with $q_1$ and the second triplet $\left(x_2, h_2, y_2 \right)$, and so on, until the end of the sequence. This construction ensures that any word $s = \left(x_1, h_1, y_1 \right),\left(x_2, h_2, y_2 \right)  \dots  \left(x_k, h_k, y_k \right)$ such that $y_k=1$ will also be accepted by the IPTA, and that $\llbracket h_k, h_k \rrbracket \sqsubseteq \Gamma(q_k)$. For an example, Figure~\ref{fig:example_ipta} shows the resulting IPTA from the set of traces 
$S$:

\[
S = \left\{
\begin{array}{rcl} 
s_1& =& (1.4, h_{1}^{1}, 1), \: (-1.07, h_{1}^{2}, 1), \: (1.08, h_{1}^{3}, 1), \: (-7.06, h_{1}^{4}, 1), \: (9.03, h_{1}^{5}, 1), \: \dots \\
s_2& =& (3.39, h_{2}^{1}, 1), \: (-3.2, h_{2}^{2}, 1),  \:(7.91, h_{2}^{3}, 1), \: (-3.45, h_{2}^{4}, 1),  \:(2.1, h_{2}^{5}, 1), \: \dots \\ 
s_3& =& (1.9, h_{3}^{1}, 1), \: (3.56, h_{3}^{2}, 1), \: (3.14, h_{3}^{3}, 0), \: (-33.2, h_{3}^{4}, 0),  \:\dots \\ s_4& =& (2.3, h_{4}^{1}, 1), \: (2.29, h_{4}^{2}, 1), \: (2.06, h_{4}^{3}, 0), \: (-0.51, h_{4}^{4}, 0) \\ \dots
\end{array}
\right.
\]

As shown in the example, there are already some merging involved when building the IPTA, since, for each state of the IPTA, there are at most $n$ outgoing transitions (one for each class of the partition). Consequently, the language accepted by the IPTA is larger than the set of traces $S$, and there is the implicit assumption that the set of traces is coherent w.r.t. the chosen partition $\Pi$. It can be checked by verifying that there cannot be two sequences $ s = \left( x_1, h_1, y_1 \right) \dots \left( x_k, h_k, y_k \right)$ and $s' = \left(x'_1, h'_1, y'_1 \right) \dots \left(x'_k, h'_k, y'_k \right)  $ such that: $s$ and $s'$ are prefixes of sequences of $S$, $\forall \: 1 \leq i  \leq k, \: \Pi^{-1}(x_i) = \Pi^{-1}(x'_i)$, and $y_k  \neq y'_k$.

If that property does not hold, it means the partition is too coarse to even build the IPTA, and that our method cannot yield a faithful representation of the behavior of the RNN. In that case, we can try again with a finer partition.

\paragraph{Merging the states} The second step of the algorithm is to merge states according to their similarity score (a real number between $0$ and $2$), as long as it is possible. A description of our method is described in algorithm \ref{alg:mergeila}.

\begin{algorithm}[!h]
\caption{Merge ILA}
\label{alg:mergeila}
\begin{algorithmic}[1]
\Input{$\mathcal{A} = \langle \Lambda, \Pi, Q, \mathbb{I}, \mathbb{F}, \delta, \Gamma \rangle$, $q_i$ and $q_j$ to be merged}
\Output{$\mathcal{A}_m = \langle \Lambda, \Pi, Q_m, \mathbb{I}_m, \mathbb{F}_m, \delta_m, \Gamma_m \rangle$ with $q_j$ merged into $q_i$}
\Initialize{\strut
$Q_m \gets Q $ \\ 
$\mathbb{I}_m \gets \mathbb{I} $ \\
$\mathbb{F}_m \gets \mathbb{F} $ \\
$\delta_m \gets \delta$ \\
$\Gamma_m \gets \Gamma$
}

\For{$q \in Q_m$}
\If{$\exists (q, \lambda, q_j) \in \delta_m$}
\State $\mathcal{A}_m \gets \Call{Add transition}{\mathcal{A}_{m}, (q, \lambda, q_i)}$
\EndIf
\If{$\exists (q_j, \lambda, q) \in \delta_m$}
\State $\mathcal{A}_m \gets \Call{Add transition}{\mathcal{A}_{m}, (q_i, \lambda, q)}$
\EndIf
\EndFor

\State $\Gamma_{m}(q_i) \gets \Gamma_{m}(q_i) \sqcup \Gamma_{m}(q_j)$
\If{$q_j \in \mathbb{I}_m$}
\State $\mathbb{I}_m \gets \mathbb{I}_m \cup \{q_i\}$;
\EndIf
\If{$q_j \in \mathbb{F}_m$}
\State $\mathbb{F}_m \gets \mathbb{F}_m \cup \{q_i\}$;
\EndIf
\State $\mathcal{A}_m \gets \Call{Delete State}{\mathcal{A}_m, q_j}$
\Return $\mathcal{A}_m$
\end{algorithmic}
\end{algorithm}

\begin{figure}[!h]
\begin{tikzpicture}[shorten >=1.0pt,node distance=3.2cm, on grid, thick,>={Stealth[round]}]

  \node[state,initial, accepting] (q_0) {\tiny $q_0 + q_2$};
  \node[state, accepting] (q_1) [above right=of q_0] {$q_1$};
  \node[state, accepting] (q_20) [right=of q_1] {$q_{20}$};
  \node[state, accepting] (q_3) [right=of q_0] {$q_3$};
  \node[state, accepting] (q_4) [right=of q_3] {$q_4$};
  \node[] (q_4+) [right=of q_4] {$\dots$};
  \node[state] (q_21) [right=of q_20] {$q_{21}$};
  \node[] (q_21+) [right=of q_21] {$\dots$};

  \path[->] (q_0) edge node [below right]{$\scalemath{0.9}{\llbracket 1.4, 3.39 \rrbracket}$} (q_1)
            (q_1) edge [bend right] node [above left]{$\scalemath{0.9}{\llbracket -3.2, -1.07 \rrbracket}$} (q_0)
            (q_0) edge node [below left, pos=0.8]{$\scalemath{0.9}{\llbracket 1.08, 7.91 \rrbracket}$} (q_3)
            (q_3) edge node [below left, pos=1.0]{$\scalemath{0.9}{\llbracket -7.06, -3.45 \rrbracket}$} (q_4)
            (q_4) edge node [below left, pos=0.8]{$\scalemath{0.9}{\llbracket 2.1, 9.03 \rrbracket}$} (q_4+)
            (q_1) edge node [above right, pos=0.05]{$\scalemath{0.9}{\llbracket 2.29, 3.56 \rrbracket}$} (q_20)
            (q_20) edge node [above left, pos=0.9]{$\scalemath{0.9}{\llbracket 2.06, 3.14 \rrbracket}$} (q_21)
            (q_21) edge node [above left, pos=0.9]{$\scalemath{0.9}{\llbracket -33.2, -0.51 \rrbracket}$} (q_21+);
\end{tikzpicture}
\caption{Result of the merging of $q_0$ and $q_2$ (from the IPTA)}
\label{fig:one_merge_step}
\end{figure}
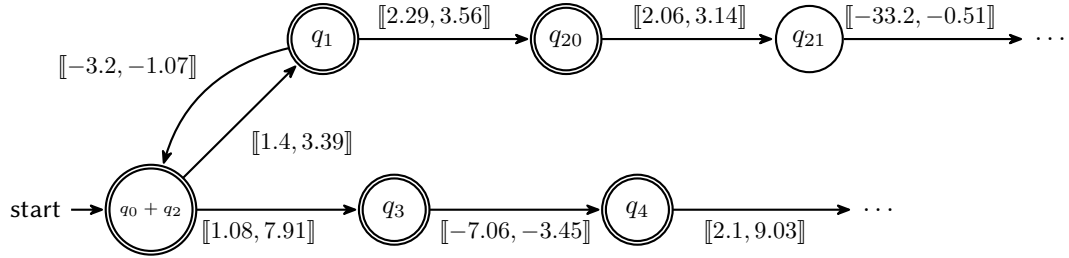

In our examples and experiments, the similarity score of two states $q_1$ and $q_2$ is defined as follows:

\begin{itemize}
\item If only one of the two states $q_1$ and $q_2$ belongs to $\mathbb{F}$, then the score is $2$
\item If both states (or none) belong(s) to $\mathbb{F}$, then the score is: 
$$ \small \hspace*{-1cm} \Call{Similarity score}{q_i, q_j} = 1 - \cos\left(mid(\Gamma(q_i)), mid(\Gamma(q_i))\right) = 1 - \frac{mid(\Gamma(q_i)) \cdot mid(\Gamma(q_j))}{\lVert mid(\Gamma(q_i)) \rVert_2 \, \lVert mid(\Gamma(q_j))\rVert_2}$$ where $mid(\Gamma(q_i))$ denotes the center of the box $\Gamma(q_i)$. 
\end{itemize}
Therefore, while there is at least one couple of states $q_1 \neq q_2$ such that their similarity score is lower than a hyperparameter $d$, the two states are merged. It means that any transition that goes to (or originates from) $q_1$ or $q_2$ will go to (or originate from) the merged state. In this process, if two transitions $(q,\lambda,q')$ and $(q,\lambda',q')$ belong to the same partition class, they will be merged. For example, when we merge the two states $q_0$ and $q_2$ of the IPTA depicted in Figure~\ref{fig:example_ipta}, we obtain the ILA depicted in Figure~\ref{fig:one_merge_step}.

At the end of the algorithm, we obtain an ILA $\mathcal{A}$ that recognizes a language larger than the set of traces $S$ given as its input. Indeed, if $ s = \left( x_1, h_1, y_1 \right) \dots \left( x_k, h_k, y_k \right)$ is a trace in $S$, then there exists a sequence of states and transitions $q_0 \xrightarrow{\lambda_1} q_1 \xrightarrow{\lambda_2} \dots \xrightarrow{\lambda_k} q_k$ in $\mathcal{A}$ such that: $q_0 \in \mathbb{I}$; $\forall \: i=1 \dots k$ we have $(q_{i-1},\lambda_i,q_i) \in \delta$, $x_i \sqsubseteq \lambda_i$, if $y_i=1$ then $q_i \in \mathbb{F}$, and $h_i \sqsubseteq \Gamma(q_i)$.

However, this ILA $\mathcal{A}$ may also accept words that are not accepted by the original RNN, and if the set of traces is too small, it may also reject words that are accepted by the original RNN. It is why we need experimental results to assess the faithfulness of the resulting automaton w.r.t the original RNN.

\section{Experiments}
\label{sec:experiments}

\paragraph{Experimental setting} Our experiments were run on a Dell Inc. Precision 3591 computer, equipped with an Intel® Core™ Ultra 7 165H × 22 CPU. Python 3.12.3 was used for algorithms implementation and data synthesis tasks. We also used PyRAT \cite{lemesle2024neural} to define the abstractions.

We evaluated our algorithm on two benchmarks. First we run experiments with RNNs trained on Tomita languages, to compare our findings with \cite{merrill2022extractingfiniteautomatarnns} and demonstrate that our approach for inferring ILA is on par with SoTA\footnote{We identified and corrected implementation errors in the lattice automaton merging and minimization functions, as well as in training the RNNs. The results reported here supersede those in the workshop version.}. 
Then, we propose a novel benchmark by extending the Tomita languages to accept sequences of floats as inputs. 
For example, the first language accepts only sequences of numbers between $0$ and $10$. The complete definition is given in Table~\ref{tab:tomita2.0_definitions}.

\begin{table}[!h]
\centering
\resizebox{\textwidth}{!}{
\begin{tabular}{|l|l|}
\hline
Tomita 2.0 language & Language description                                                                   \\ \hline
1                   & ($\llbracket 0, 10 \rrbracket)^{\star}$                                                          \\ \hline
2                   & $(\llbracket 0, 10 \rrbracket \llbracket -10, 0 \llbracket)^{\star}$                    \\ \hline
3                   & No odd $\llbracket -10, 0 \llbracket$ after an odd $\llbracket 0, 10 \rrbracket$ string \\ \hline
4                   & No substring containing 3 consecutive letters $\in \llbracket -10, 0 \llbracket$        \\ \hline
5 & Even number of letters $\in \llbracket -10, 0 \llbracket$ and even number of letters $\in \llbracket 0, 10 \rrbracket$                           \\ \hline
6 & Difference of the number of letters $\in \llbracket 0, 10 \rrbracket$ and letters $\in \llbracket -10, 0 \llbracket$ is a multiple of 3         \\ \hline
7 & $\llbracket -10, 0 \llbracket^{\star} \llbracket 0, 10 \rrbracket^{\star}\llbracket -10, 0 \llbracket^{\star} \llbracket 0, 10 \rrbracket^{\star}$ \\ \hline
\end{tabular}
}
\caption{Definitions of the 7 Tomita 2.0 languages used in the experiments}
\label{tab:tomita2.0_definitions}
\end{table}

Tables \ref{tab:benchs1} and \ref{tab:benchs2} summarize our results. Our implementation takes roughly few seconds to infer an automaton from a set of execution traces. The fidelity score is measured as follows $\Call{Fidelity}{\mathcal{R}, \mathcal{A}, x} = \sum_{i=1}^{n} |\mathcal{R}(x_i) - \mathcal{A}(x_i)|$ for a given RNN $\mathcal{R}$, ILA $\mathcal{A}$ and sequence of inputs $x$. 

\begin{figure}[!h]
    \centering
    \includegraphics[width=0.9\columnwidth]{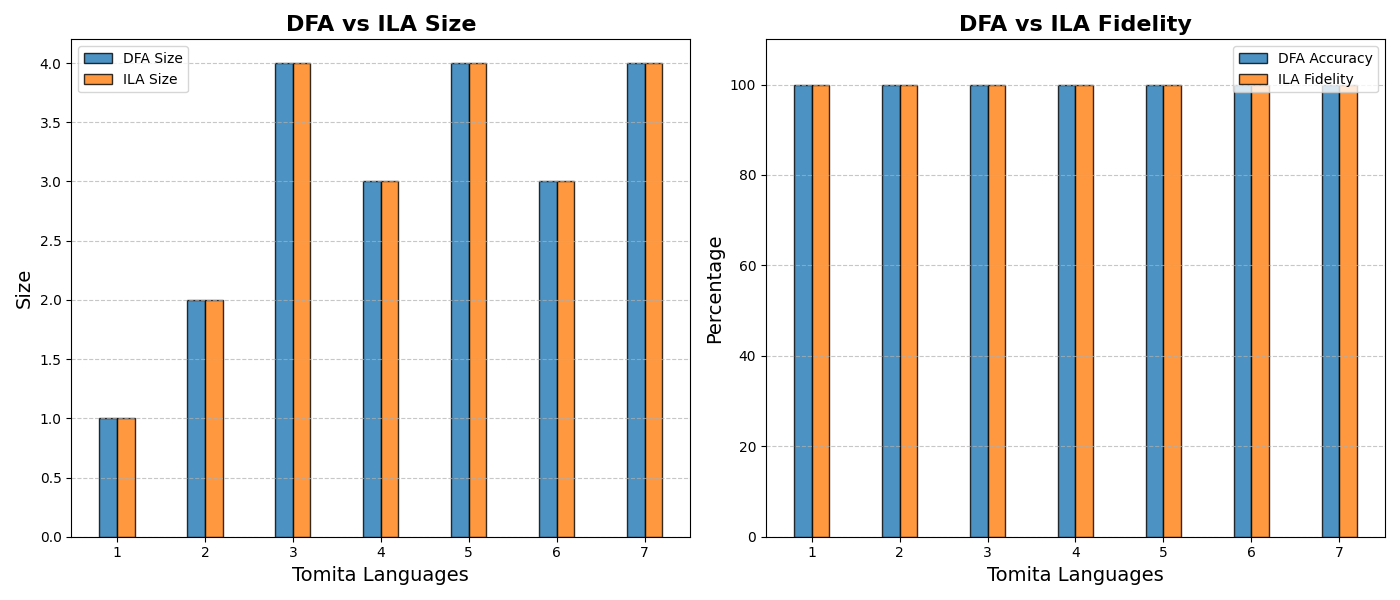}
    \caption{Fidelity and size comparison between ILA and DFA}
    \label{fig:results_plot}
\end{figure}

\begin{table}[!h]
\centering
\begin{tabular}{| p{3.75cm} | p{0.85cm} p{0.85cm} p{0.85cm} p{0.85cm} p{0.85cm} p{0.85cm} p{0.85cm} |}
\hline
\textbf{Tomita language} & \textbf{1}& \textbf{2}& \textbf{3}& \textbf{4}& \textbf{5}& \textbf{6}& \textbf{7}\\ \hline \hline
RNN accuracy     & 100 & 100   & 100 & 100 & 100 & 99.89 & 100   \\ \hline
ILA size         & 1   & 2     & 4  & 3   & 4  & 3    & 4    \\ \hline 
ILA Fidelity     & 100 & 100 & 100 & 100 & 100 & 100   & 100 \\ \hline
Type I error     & 0   & 0  & 0   & 0   & 0   & 0    & 0 \\ \hline
Type II error    & 0   & 0 & 0   & 0   & 0   & 0     & 0  \\ \hline
\end{tabular}
\caption{Benchmark results of passive learning of ILA from RNNs trained on Tomita languages}
\label{tab:benchs1}
\end{table}

\begin{table}[!h]
\centering
\begin{tabular}{| p{3.75cm} | p{0.85cm} p{0.85cm} p{0.85cm} p{0.85cm} p{0.85cm} p{0.85cm} p{0.85cm} |}
\hline
\textbf{Tomita 2.0 language} & \textbf{1}& \textbf{2}& \textbf{3}& \textbf{4}& \textbf{5}& \textbf{6}& \textbf{7}     \\ \hline \hline
RNN accuracy        & 100 & 100 & 100 & 100 & 100 & 100 & 100 \\ \hline
ILA size            & 1    & 2     & 4    & 3    & 4    & 3     & 4    \\ \hline
ILA Fidelity        & 100 & 100 & 100 & 100 & 100 & 100 & 100 \\ \hline
Type I error        & 0  & 0  & 0  & 0  & 0   & 0  & 0     \\ \hline
Type II error       & 0 & 0  & 0 & 0  & 0 & 0 & 0 \\ \hline
\end{tabular}
\caption{Benchmark results of passive learning for ILA from RNNs trained on Tomita 2.0 languages}
\label{tab:benchs2}
\end{table}

We also analyze type I and type II errors to evaluate our algorithm:
\begin{itemize}
\item A type I error would occur because of abstractions imprecision, when a word is rejected by the RNN and accepted by its automaton (also known as false alarms).
\item A type II error reflects failure to capture RNN semantics, which can be due to insufficient sample size used to build the IPTA. 
\end{itemize}

The scores of extracted ILA from a Tomita 2.0 RNN are similar to state of the art compared to Tomita languages. It is trivial to see that the automata topologies and number of states for Tomita 2.0 is exactly the same as for Tomita. Our preliminary results can be improved by applying them for RNNs trained on real world tasks experimenting with different state merging algorithms that perform better than Gold's algorithm, such as RPNI or EDSM \cite{Soubki2023BenchmarkingSA}.

\section{Conclusion and Future Work}
\label{sec:discussion}
We presented in this paper a passive learning algorithm that is capable of inferring an automaton from a set of traces of an RNN. Unlike previous methods, we do not require the inputs of the RNN to belong to a finite alphabet. Our algorithm ensures that we obtain an overapproximation of the set of traces. 
Our experiments demonstrate a capacity to efficiently infer ILA from an RNN trained on possibly infinite regular language.

In future work, we aim to extend this proof of concept, especially for the verification and the explainability of RNNs trained for practical real-world applications, \eg time series forecasting, where the robustness of the RNN is of paramount importance. The automata-based approach is a generalizable formal method offering a synergy between the interpretability and the verification for recurrent networks. While properties that can be formulated and verified are beyond adversarial robustness, the explanation can also be regarded as a property we seek to verify.

\paragraph{Acknowledgements}: This work was supported by the French Agence Nationale de la Recherche (ANR) through SAIF (ANR-23-PEIA-0006) as part of the France 2030 programme.

\newpage
\bibliography{ref}

@String{Springer = "Springer-Verlag" }

@misc{lin2023segrnnsegmentrecurrentneural,
  title={Segrnn: Segment recurrent neural network for long-term time series forecasting},
  author={Lin, Shengsheng and Lin, Weiwei and Wu, Wentai and Zhao, Feiyu and Mo, Ruichao and Zhang, Haotong},
  journal={arXiv preprint arXiv:2308.11200},
  year={2023}
}

@article{Schrittwieser_2020,
  title={Mastering atari, go, chess and shogi by planning with a learned model},
  author={Schrittwieser, Julian and Antonoglou, Ioannis and Hubert, Thomas and Simonyan, Karen and Sifre, Laurent and Schmitt, Simon and Guez, Arthur and Lockhart, Edward and Hassabis, Demis and Graepel, Thore and others},
  journal={Nature},
  volume={588},
  number={7839},
  pages={604--609},
  year={2020},
  publisher={Nature Publishing Group UK London}
}

@misc{merrill2022extractingfiniteautomatarnns,
  title={Extracting finite automata from RNNs using state merging},
  author={Merrill, William and Tsilivis, Nikolaos},
  journal={arXiv preprint arXiv:2201.12451},
  year={2022}
}

@article{adax,
  title={Adaax: Explaining recurrent neural networks by learning automata with adaptive states},
  author={Hong, Dat and Segre, Alberto Maria and Wang, Tong},
  booktitle={Proceedings of the 28th ACM SIGKDD Conference on Knowledge Discovery and Data Mining},
  pages={574--584},
  year={2022}
}

@article{Mayr2021PropertyCW,
  title={Property checking with interpretable error characterization for recurrent neural networks},
  author={Mayr, Franz and Yovine, Sergio and Visca, Ramiro},
  journal={Machine Learning and Knowledge Extraction},
  volume={3},
  number={1},
  pages={205--227},
  year={2021},
  publisher={MDPI}
}

@misc{muškardin2023relationshiprnnhiddenstate,
  title={On the Relationship Between RNN Hidden State Vectors and Semantic Ground Truth},
  author={Mu{\v{s}}kardin, Edi and Tappler, Martin and Pill, Ingo and Aichernig, Bernhard K and Pock, Thomas},
  journal={arXiv preprint arXiv:2306.16854},
  year={2023}
}

@article{ANGLUIN198787,
  title={Learning regular sets from queries and counterexamples},
  author={Angluin, Dana},
  journal={Information and computation},
  volume={75},
  number={2},
  pages={87--106},
  year={1987},
  publisher={Elsevier}
}

@article{Jacobsson2005RuleEF,
  title={Rule extraction from recurrent neural networks: Ataxonomy and review},
  author={Jacobsson, Henrik},
  journal={Neural Computation},
  volume={17},
  number={6},
  pages={1223--1263},
  year={2005},
  publisher={MIT Press One Rogers Street, Cambridge, MA 02142-1209, USA journals-info~…}
}

@inproceedings{Bollig2022ASO,
  title={A survey of model learning techniques for recurrent neural networks},
  author={Bollig, Benedikt and Leucker, Martin and Neider, Daniel},
  journal={A Journey from Process Algebra via Timed Automata to Model Learning: Essays Dedicated to Frits Vaandrager on the Occasion of His 60th Birthday},
  pages={81--97},
  year={2022},
  publisher={Springer}
}

@inproceedings{Gall2007LatticeAA,
  title={Lattice automata: A representation for languages on infinite alphabets, and some applications to verification},
  author={Le Gall, Tristan and Jeannet, Bertrand},
  booktitle={International Static Analysis Symposium},
  pages={52--68},
  year={2007},
  organization={Springer}
}

@article{GOLD1967447,
  title={Language identification in the limit},
  author={Gold, E Mark},
  journal={Information and control},
  volume={10},
  number={5},
  pages={447--474},
  year={1967},
  publisher={Elsevier}
}

@inproceedings{Mukardin2022LearningFS,
  title={Learning finite state models from recurrent neural networks},
  author={Mu{\v{s}}kardin, Edi and Aichernig, Bernhard K and Pill, Ingo and Tappler, Martin},
  booktitle={International Conference on Integrated Formal Methods},
  pages={229--248},
  year={2022},
  organization={Springer}
}

@article{Wei2023WeightedAE,
  title={Weighted automata extraction and explanation of recurrent neural networks for natural language tasks},
  author={Wei, Zeming and Zhang, Xiyue and Zhang, Yihao and Sun, Meng},
  journal={Journal of Logical and Algebraic Methods in Programming},
  volume={136},
  pages={100907},
  year={2024},
  publisher={Elsevier}
}

@article{Song2023LUNAAM,
  title={Luna: A model-based universal analysis framework for large language models},
  author={Song, Da and Xie, Xuan and Song, Jiayang and Zhu, Derui and Huang, Yuheng and Juefei-Xu, Felix and Ma, Lei},
  journal={IEEE Transactions on Software Engineering},
  volume={50},
  number={7},
  pages={1921--1948},
  year={2024},
  publisher={IEEE}
}

@article{Dong2019AnalyzingRN,
  title={Towards interpreting recurrent neural networks through probabilistic abstraction},
  author={Dong, Guoliang and Wang, Jingyi and Sun, Jun and Zhang, Yang and Wang, Xinyu and Dai, Ting and Dong, Jin Song and Wang, Xingen},
  booktitle={Proceedings of the 35th IEEE/ACM International Conference on Automated Software Engineering},
  pages={499--510},
  year={2020}
}

@article{Wang2018ACS,
  title={A comparative study of rule extraction for recurrent neural networks},
  author={Wang, Qinglong and Zhang, Kaixuan and Ororbia II, Alexander G and Xing, Xinyu and Liu, Xue and Giles, C Lee},
  journal={arXiv preprint arXiv:1801.05420},
  year={2018}
}

@inproceedings{CC77,
  title={Abstract interpretation: a unified lattice model for static analysis of programs by construction or approximation of fixpoints},
  author={Cousot, Patrick and Cousot, Radhia},
  booktitle={Proceedings of the 4th ACM SIGACT-SIGPLAN symposium on Principles of programming languages},
  pages={238--252},
  year={1977}
}

@inproceedings{Johnson2016SupervisedAS,
  title={Supervised and semi-supervised text categorization using LSTM for region embeddings},
  author={Johnson, Rie and Zhang, Tong},
  booktitle={International Conference on Machine Learning},
  pages={526--534},
  year={2016},
  organization={PMLR}
}

@inproceedings{Soubki2023BenchmarkingSA,
  title={Benchmarking State-Merging Algorithms for Learning Regular Languages},
  author={Soubki, Adil and Heinz, Jeffrey},
  booktitle={International Conference on Grammatical Inference},
  pages={181--198},
  year={2023},
  organization={PMLR}
}

@article{lemesle2024neural,
  title={Neural network verification with pyrat},
  author={Lemesle, Augustin and Lehmann, Julien and Gall, Tristan Le},
  journal={arXiv preprint arXiv:2410.23903},
  year={2024}
}
\end{document}